\documentclass[aps,prl,preprint,groupedaddress]{revtex4-1}
\usepackage{graphicx}
\usepackage{epsfig,amsmath,subfigure}
\usepackage{filecontents}
\usepackage{setspace}

\newcommand{\dd}{\textrm{d}}
\newcommand{\pd}{\partial}

\begin{document}

%\preprint{}

%\title{Evolution of Faraday waves by resonant triad interactions of surface--compression waves}
\title{Faraday waves by resonant triad interaction of surface--compression waves}


\author{Usama Kadri}
%\thanks{}
\affiliation{School of Mathematics, Cardiff University, Cardiff, CF24 4AG, UK}
\affiliation{Department of Mathematics, Massachusetts Institute of Technology, Cambridge, MA 02139, USA}


\date{\today}

\begin{abstract}
The propagation of wave disturbances over a vertically oscillating liquid may form standing waves, known as Faraday waves. Here we present an alternative description of the generation and evolution of Faraday waves by nonlinear resonant triad interactions, accounting for gravity surface effects and the slight compressibility of the liquid. To this end, we analyse a triad comprising an infinitely long-crested compression mode and two oppositely propagating subharmonic surface waves.
\end{abstract}

%\keywords{}

\maketitle
\section{Introduction}
A surface wave disturbance over a fluid layer that is subject to sinusoidal vertical oscillation at frequency $2\omega$ may excite subharmonic standing field of waves of frequency $\omega$ known as Faraday waves. The wavelength of these waves is prescribed by the standard dispersion relation \cite{benjamin1954stability}. For many applications the driven flow due to the interaction of the disturbance with the oscillating bath is attributed to parametric instability (e.g. see Ref. \cite{milewski2015faraday}) (or parametric resonance). Here, we introduce an alternative approach to describe this energy share between the surface waves and the oscillating bath. We model the oscillating bath as an infinitely long-crested compression (acoustic) wave that exchanges energy with surface (gravity) waves via resonant triad interactions.  

Special triad interactions between surface wave disturbances and compression modes associated with resonances were noted by Longuet-Higgins \cite{longuet1950theory}, and later by Kadri \& Stiassnie \cite{kadri2013generation}. More specifically, these comprise a standing compression mode at a cut-off frequency $2\omega$, and two subharmonic surface waves with frequency $\omega$ and opposite wavenumbers $k$. Here, we consider the interaction of a propagating radial surface wave disturbances and a standing compression mode in a deep fluid. Such resonance interaction depends on the small parameter $\mu = {gh}/{c^2}$, $(\mu\ll 1)$ (see Ref. \cite{kadri2016resonant}) that governs the effects of gravity relative to compressibility, where $g$ is the gravitational acceleration. Thus, free-surface wave disturbances feature vastly different spatial and/or temporal scales from the compression mode. Specifically, the surface wavelength $\lambda$ is much shorter than the vertical compression lengthscale represented by the water depth $(\lambda\ll h)$, whereas the gravity timescale $\tau\sim (\lambda/g)^{1/2}$ in keeping with the deep-water dispersion relation. Then, taking $\tau$ to be comparable to the acoustic timescale $h/c$ implies $\lambda\sim \mu h$; hence, in the present setting, the parameter $\mu$ may be interpreted as the ratio of the gravity to the (vertical) compression mode lengthscale.

\section{Preliminaries}
Based on irrotationality, the surface--compression wave problem is formulated in terms of the velocity potential $\varphi(r,z,t)$, where $\mathbf{u}=\nabla\varphi$ is the velocity field. Moreover, we shall use dimensionless variables, employing $\mu h$ as lengthscale and $h/c$ as timescale. We shall also assume radial motion ($\partial_{\theta}=0$), far from the centre ($1/r\ll 1$); then the field equation obeys
\begin{equation} \label{eq:field}
\varphi_{tt}-\frac{1}{\mu^2}\left(\varphi_{rr}+\varphi_{zz}\right)+\varphi_z+|\nabla\varphi|^2_t+\tfrac{1}{2}\mathbf{u}\cdot \nabla\left(|\nabla\varphi|^2\right)=0.
\end{equation}
On the free surface $z=\eta(r,t)$, we consider the standard dynamic and kinematic conditions. Finally, the boundary condition on the rigid bottom at $z=-1/\mu$ reads
\begin{equation} \label{eq:bc_b}
\varphi_{z}=0 \qquad (z=-1/\mu).
\end{equation}

The velocity potential for the three interacting modes is expanded as follows
\begin{equation} \label{eq:triad_potential}
\begin{split}
\varphi=\epsilon~\textrm{e}^{kz}\Big\{S_+(T)\textrm{e}^{\textrm{i}(kr-\omega t)}+ S_-(T)\textrm{e}^{-\textrm{i}(kr+\omega t)}+\textrm{c.c.}\Big\}
+\alpha\cos{\omega(Z+1)}\Big\{C(T)\textrm{e}^{-2\textrm{i} \omega t}+\textrm{c.c.}\Big\}+\dots.\\
\end{split}
\end{equation}
%where $\Theta_{\pm}=k_{\pm}x-\omega_{\pm}t$ and $\Theta=\mu\kappa x-\omega t$.
The first terms in (\ref{eq:triad_potential}) represent the two surface waves while the last represents the compression mode, that is scaled in the vertical coordinate $Z=\mu z$. The surface wave amplitudes $S_{\pm}$ and the compression mode amplitude $C$ depend on the `slow' time $T=\mu t$, and space $R=\mu r$, where $\epsilon=\alpha\mu^{1/2}$ with $\alpha=O(1)$  (see Ref. \cite{kadri2016resonant}). 

\section{Evolution equations}
Upon substituting (\ref{eq:triad_potential}) in the surface--compression mode problem (\ref{eq:field}) and the boundary conditions, we focus on terms proportional to $\exp\{\textrm{i}(kr-\omega t)\}$, $\exp\{-\textrm{i}(kr+\omega t)\}$, and $\exp\{-2\textrm{i}\omega t\}$. These terms cause secular behaviour at higher order in expansion (\ref{eq:triad_potential}) as they have the same spatial and temporal dependence as the three linear propagation modes at the leading order. To overcome this difficulty we impose solvability conditions on the problems governing higher-order corrections to these modes, which results in the desired evolution equations for the compression wave amplitude $C(T)$ 
\begin{equation} \label{eq:A_c}
\frac{\pd C}{\pd T}=-\frac{\textrm{i}}{2\omega}\frac{\pd^2 C}{\pd R^2}-\frac{{\textrm{i}}}{2\omega} C+\frac{1}{4}\omega^3\alpha S_+S_-.
%\textrm{i}{\pd_\tau C}=\frac{\textrm{1}}{2\omega}{\pd_{\xi\xi}^2 C}+\frac{{\textrm{1}}}{2\omega} C+\frac{1}{4}\omega^3\alpha S_+S_-.
\end{equation}
and the surface wave amplitudes $S_{\pm}(T)$
\begin{equation} \label{eq:S_b}
\frac{\dd S_{\pm}}{\dd T}=-\frac{1}{8}\omega^3\alpha CS_{\mp}^*-\frac{\textrm{i}}{64}\omega^7\alpha^2\left(S_{\pm}^2S_{\pm}^*-4|S_{\mp}|^2S_{\pm}\right),
\end{equation}
where $*$ stands for complex conjugate. Note that equation (\ref{eq:S_b}) is a particular case of equation (5.9) derived by Kadri \& Akylas \cite{kadri2016resonant}.

\section{Results and discussion}
A qualitative comparison between experiments (originally presented in \cite{eddi2011information}) and the proposed mechanism are shown in figure 1(a)-(d). The amplitude evolution of a surface wave is given in figure 1(e). Note that in the given example, the prescribed surface waves are travelling in opposite directions and thus present a standing disturbance in analogy to a disturbance by a falling steel ball impacting the surface. Unlike parametric resonance, with the proposed mechanism we can also consider a moving disturbance, say due to a moving droplet (e.g. see \cite{milewski2015faraday}), resulting in a rigorous description of the evolution of the disturbance path memory.
\begin{figure} [h]
\centering{ \epsfig{figure=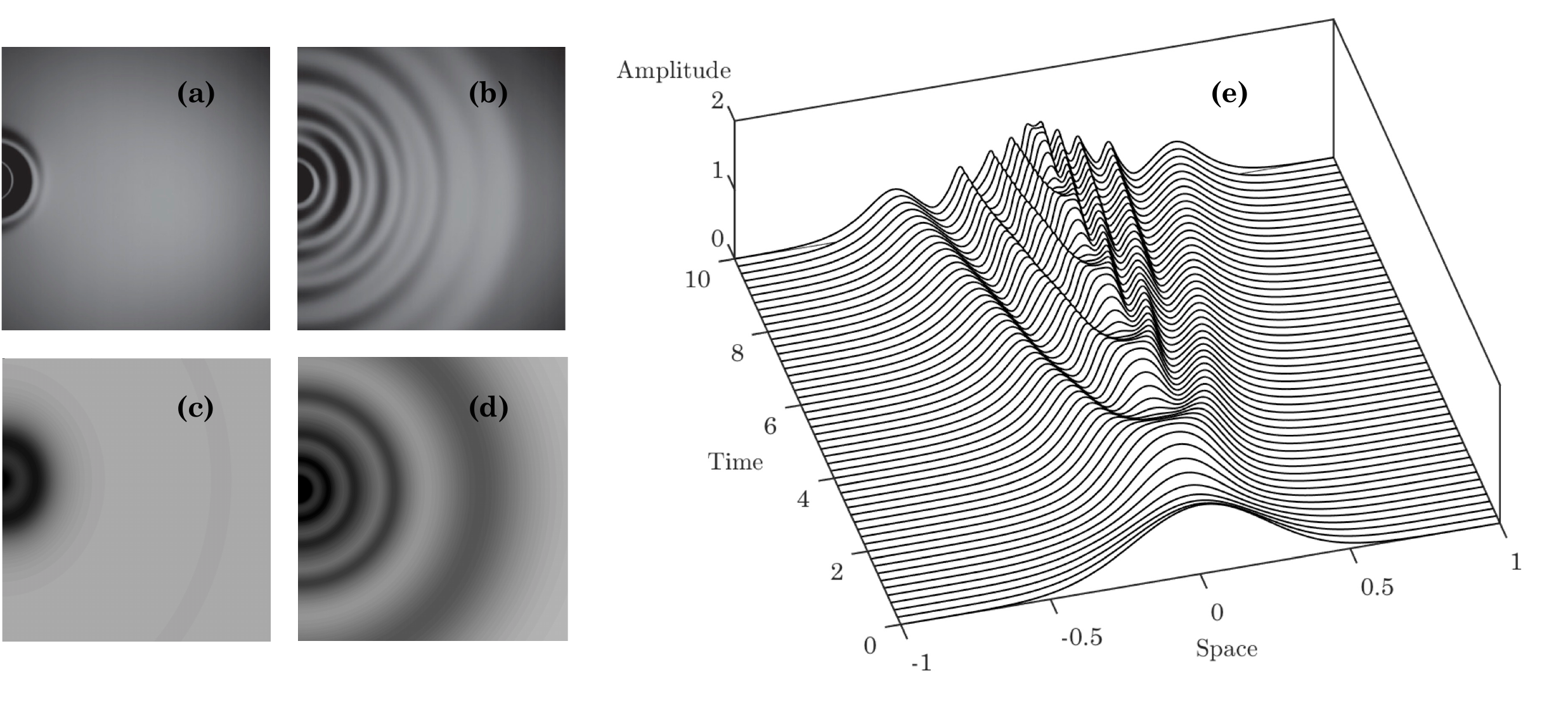,width=15cm}}
\caption{(a,b): Photographs of the wave field generated by a falling steel ball on a vertically oscillating liquid bath, at times t=51 ms (a) and t=173 ms (b) after the collision --- (a,b) are presented in \cite{eddi2011information}. (c,d): Simulations of the wave field generated from nonlinear triad resonance of two surface waves propagating in opposite directions, and a compression mode oscillating in the vertical direction, at times $T=17$ (c) and $T=51$ (d). (e): Evolution of the surface wave amplitude during the resonant interaction.
}  \label{fig:sim_exp}
\end{figure}

\end{document}